\documentclass[11pt]{article}

\usepackage{lineno}
\usepackage{hyperref}
\usepackage{amssymb}

\modulolinenumbers[5]

\begin{document}

\newtheorem{theorem}{Theorem}[section]
\newtheorem{lemma}[theorem]{Lemma}
\newtheorem{corollary}[theorem]{Corollary}
\newtheorem{proposition}[theorem]{Proposition}
\newcommand{\blackslug}{\penalty 1000\hbox{
    \vrule height 8pt width .4pt\hskip -.4pt
    \vbox{\hrule width 8pt height .4pt\vskip -.4pt
          \vskip 8pt
      \vskip -.4pt\hrule width 8pt height .4pt}
    \hskip -3.9pt
    \vrule height 8pt width .4pt}}
\newcommand{\proofend}{\quad\blackslug}
\newenvironment{proof}{$\;$\newline \noindent {Proof.}$\;$\rm}{\qad}
\newenvironment{correct}{$\;$\newline \noindent {\sc Correctness.}$\;\;\;$\rm}{\qad}
\newcommand{\qad}{\hspace*{\fill}\blackslug}
\newenvironment{definition}{$\;$\newline \noindent {\bf Definition}$\;$}{$\;$\newline}
\def\boxit#1{\vbox{\hrule\hbox{\vrule\kern4pt
  \vbox{\kern1pt#1\kern1pt}
\kern2pt\vrule}\hrule}}
\def\ord{\mbox{\it ord}}
\def\maf{\mbox{\bf maf }}
\def\lca{\mbox{\it lca}}


\title{\vspace{-.5in}
{\bf Linear-Time Parameterized Algorithms\\ with Limited Local Resources}\thanks{This 
work is supported in part by the National Natural Science Foundation of China under grants 
61420106009 and 61872097.}}

\author{{\sc Jianer Chen}$^{\dagger\ddagger}$, {\sc Ying Guo}$^\dagger$, and {\sc Qin Huang}$^\ddagger$
\vspace*{3mm}\\
$^\dagger$ School of Computer Science\\
    Guangzhou University,
    Guangzhou 510006, P.R.~China
\vspace*{1mm}\\
$^\ddagger$ Department of Computer Science and Engineering \\
    Texas A\&M University,
    College Station,  TX 77843, USA
}

\date{}

\maketitle

\begin{abstract} 
We propose a new (theoretical) computational model for the study of massive data processing with limited 
computational resources. Our model measures the complexity of reading the very large data sets in terms of 
the data size $N$ and analyzes the computational cost in terms of a parameter $k$ that characterizes the 
computational power provided by limited local computing resources. We develop new algorithmic techniques that 
implement algorithms for solving well-known computational problems on the proposed model. In particular, we 
present an algorithm that finds a $k$-matching in a general unweighted graph in time $O(N + k^{2.5})$ and 
an algorithm that constructs a maximum weighted $k$-matching in a general weighted graph in time 
$O(N + k^3 \log k)$. Both algorithms have their space complexity bounded by $O(k^2)$. 

\vspace{2mm}

\noindent {\bf keywords.}\  bigdata, linear-time algorithm, space complexity, graph matching

\end{abstract} 


\section{Motivations}

Recent progress in data science has shown that classical algorithmic techniques may become inadequate when 
dealing with data sets of enormous size. For example, Facebook has billions of users and trillions of links 
\cite{2017-18}. Thus, a traditionally ``efficient'' algorithm of running time $O(n^2)$ may turn out to be not 
practically feasible. There have been fast growing interests in the study of massive data sets. The research 
has included the study of structures of massive data and data queries (e.g., \cite{2017-14}), parallel and 
distributed processing of massive data (e.g., \cite{2017-12}), and preprocessing of massive data (e.g., 
\cite{2017-17}). The research has been driven directly by practical applications in massive data processing, 
and is essentially heuristic-based. There has also been very active research in the algorithmic community. 
The study of very fast (sublinear-time, linear-time, or nearly linear-time) algorithms in dealing with massive data 
sets has drawn extensive attention. A number of computation models for dealing with massive data sets have been 
proposed and studied. In particular, data {\it streaming} and {\it semi-streaming} models \cite{2017-2,semi} 
have been proposed and studied, where the massive data (e.g., ``big graphs'') may dynamically change and 
the algorithms must process the input stream in the order it arrives while using only a limited amount of memory. 
Very recently, studies on streaming algorithms based on the framework of parameterized computation have 
appeared \cite{2017-4,soda2016,2017-3}.

In the current paper, we propose a new (theoretical) computational model for the study of massive data processing 
with limited ``local'' computing resources. Our model is of a multivariate nature, which measures the complexity of 
reading the very large data sets in terms of the size of the data sets and analyzes the computational cost in terms 
of a parameter that characterizes the computational power provided by limited local computing resources. In particular, 
problems in our consideration have two parameters $N$ and $k$, where $N$ is the input size, which is assumed to 
be extremely large thus superlinear-time (such as quadratic-time) algorithms would be considered impractical, while 
$k$ gives the ``size'' of feasibility such that the limited local computing resource (e.g., a normal computer) can handle 
problems with complexity (time and space) bounded by a polynomial of $k$, in addition to the linear-time reading from 
the input data. More specifically, we will study algorithms for processing massive data sets that run in {\it linear-time} 
in terms of the input size $N$, and polynomial time and polynomial space in terms of the parameter $k$, i.e., 
algorithms running in time $O(N + k^{O(1)})$ and space $O(k^{O(1)})$. 

We argue that the proposed model is theoretically interesting and practically meaningful. Insisting on strict linear-time 
in terms of the size of input data sets allows us to process data sets of very large size. On the other hand, there seems 
no simple functional relations between the size of input data sets and the power of available computational resources.
In many cases, problems in massive data processing (such aggregations) on very large data sets are looking for 
solutions of size manageable by local computational resources, where the size of solutions and the size of input 
data sets do not seem to be directly correlated. Therefore, it is meaningful and convenient by introducing another 
parameter $k$ to characterize the available computational resources. The constraint on the space complexity in terms 
of the parameter $k$ reflects the fact that although massive data sets are stored publicly, users can only read the data 
but do not own the space for storing the data. Allowing the cost of local resources to be bounded by polynomials 
of the parameter $k$ offers new challenges in algorithmic research. We point out that optimizing the 
cost of local resources in terms of the parameter $k$ implies widening the applicability of the algorithms. For example, 
if $k$ is the solution size, then algorithms whose resources are bounded by lower-degree polynomials of $k$ allow us 
to handle massive data problems with larger solutions. 

As examples, we consider a number of well-known problems that have been extensively studied in algorithmic 
research, and demonstrate how these problems can be solved in the proposed model. In particular, we show how 
the famous graph matching problems (on bipartite/general and unweighted/weighted graphs) can be  solved on this 
model. We present an algorithm that finds a $k$-matching in a general unweighted graph in time $O(N + k^{2.5})$ and 
an algorithm that constructs a maximum weighted $k$-matching in a general weighted graph in time $O(N + k^3 \log k)$.

\section{Definitions and related work}

Let $A$ be an algorithm that solves a computational problem $Q$. Inputs to the algorithm $A$ take the form of pairs 
$(x, k)$, where $x$ is a proper encoding of an instance of $Q$ and $k$ is a parameter. For example, inputs to an 
algorithm that solves the {\it Maximum Weighted $k$-Matching} problem ({\sc MaxW $k$-Matching)} are of the form 
$(G, k)$, where $G$ is a weighted graph given in an adjacency list, encoded properly, and the instance is looking for a 
$k$-matching in $G$ that has the maximum weight over all $k$-matchings in the graph $G$. 

We assume that our algorithms run on the word-RAM model, in which each basic operation (e.g., arithmetic operations 
and comparison) on words (i.e., the basic elements in a problem instance) takes constant time. Moreover, we assume 
that the instances are ``word addressable'' so that algorithms can read any word in an input instance in constant time. 
On the other hand, we do not allow algorithms to write (i.e., to modify) on input data. We will be focused on algorithms 
whose running time is bounded by $O(N + k^{O(1)})$ and whose space is bounded by $O(k^{O(1)})$, measured in word 
complexity, where $N$ is the ``size'' of the input, i.e., the number of words in the input instance, and $k$ is a parameter 
independent of the input size $N$ that measures the ``local complexity'' of the algorithms. We remark that by definition, 
our algorithms will run in linear-time in the size of the input data for {\it both} word complexity and bit complexity. In fact, 
under common assumptions, the $N$ words in the input can be given in $O(N \log N)$ bits. Since each basic word 
operation takes constant time in word complexity, which is $O(\log N)$-time in bit complexity, the $O(N)$-time word 
complexity of the algorithms implies $O(N \log N)$-time bit complexity, which is linear in terms of bit complexity of the 
input data. On the other hand, the word complexity $O(k^{O(1)})$ in local time and space would have an additional $\log N$ 
factor if we use bit complexity. We remark that, unlike some other proposed models (e.g. \cite{2017-4,2017-2}), the 
complexity bounds given for an algorithm in our model are not allowed to have an ``implicit'' polylog (i.e., a polynomial 
of $\log N$ or $\log k$) factor. Therefore, the time complexity of an algorithm in our model counts up the number of 
computational steps of the algorithm.

There have been several computational models in the literature that are related to our model. 

A well-known complexity class $SC$ (Steve's Class) that bounds both time and space complexities simultaneously 
was proposed by Stephen Cook \cite{cook}, which consists of problems that are solvable in polynomial time while,  
{\it simultaneously}, with the space being bounded by $O(\log^{O(1)} n)$. In particular, the set of deterministic 
context-free languages is in the class $SC$ \cite{cook}. Because the model allows high-degree polynomials in its 
running time, it may not be suitable for the paradigm  of massive data processing. 

Motivated by massive data processing, sublinear-time algorithms have been studied recently \cite{rubinfeld}, which use 
randomization and inspect only a portion of the input data to give (in some sense imprecise) solutions. Quality of 
sublinear-time algorithms are measured in terms of the input size $N$ and an error bound $\epsilon$. On the other 
hand, linear time and quasi-linear time algorithms have been the focus in algorithmic research for years and have been 
studied extensively \cite{clrs}, where however simultaneous bound on space complexity was seldom considered.  

In the study of parameterized computation, there have been recent interests in ``linear-time kenelization'' algorithms 
\cite{rolf}. A kernelization algorithm for a parameterized problem $Q$ translates an instance $(x, k)$ of $Q$ into an 
``equivalent'' instance $(x', k')$ of $Q$ such that both $|x'|$ and $k'$ are bounded by a function of $k$. In particular, 
linear-time kernelization algorithms for maximum matching in unweighted graphs have been developed in terms 
of various parameters of the input graph, such as the feedback edge number, the feedback vertex number, and the 
distance to chain graphs \cite{rolf}. However, the kernelization algorithms given in \cite{rolf}, as well as in other recent 
work in this direction, operate on the input graph. As a result, the working space of the algorithms is a function of the 
size $N$ of the input. There has also been recent research on parameterized and kernelization algorithms for NP-hard 
problems on dynamic inputs whose major concern is on bounding the update time by $f(k) N^{1 + o(1)}$ for a 
function $f(k)$ of $k$ \cite{alman}.    

The study of streaming and semi-streaming algorithms have attracted much attention in recent years 
\cite{2017-2,stream,semi}, where the input data are given as a stream of data while the algorithms must process 
the input data in the order they arrive, within a given space bound. For instance, streaming/semi-streaming graph 
algorithms in general are restricted to space bound $O(n \log^{O(1)} n)$, where $n$ is the number of vertices on the 
input graph (thus can be sublinear in terms of the size of the graph). Another complexity measure in streaming 
algorithms is the {\it per-element process-time} or {\it update time} \cite{stream}, which, when multiplied with the 
number of elements in the input, gives an upper bound on the running time of the algorithms. 

Streaming algorithms have been studied under the framework of parameterized computation recently. Fafianie and 
Kratsch \cite{streamkernel} considered polynomial-time kernelization algorithms for streaming graphs on a number of 
NP-hard problems, where the algorithms are restricted to have space bounded by $O(k^{O(1)})$. Parameterized 
streaming algorithms have also been studied  \cite{soda2016,2017-4,2017-3}, where the focus on the streaming 
algorithms includes space bound, update time, and solution extraction time \cite{2017-3}. In particular, graph 
matching problems on the dynamic streaming model (in which the stream consists of both edge insertion and 
edge deletion operations) have been studied. Streaming algorithms of $O(k^2 \log^{O(1)} N)$ space have been 
developed for the {\it maximal} matching problem on the dynamic streaming model, under the promise that no matching 
has more than $k$ edges in any graph formed by a prefix of the stream \cite{2017-4,2017-3}. Algorithms for constructing 
a maximum matching of at most $k$ edges in an unweighted or a weighted graph have also been studied under the 
dynamic streaming model \cite{soda2016}.

\section{Case study I: matching in unweighted graphs}

In this and the next sections, we provide thorough investigations on algorithms in our proposed model that solve 
the famous graph matching problems. This section is focused on unweighted graphs, while the next section is on 
weighted graphs. 

All graphs in our discussion are undirected, which are given in the adjacency list format. A graph $G$ is {\it weighted} 
if each edge in $G$ is associated with a {\it weight}, which is a real number.

A {\it matching $M$} in a graph $G$ is a set of edges in $G$ such that no two edges in $M$ share a common 
end. A matching is a {\it $k$-matching} if it consists of exactly $k$ edges. A vertex $v$ is {\it covered} by the 
matching $M$ if $v$ is an end of an edge in $M$. Otherwise, the vertex $v$ is {\it uncovered}.

The instances of the (parameterized) {\sc Unweighted Graph Matching} problem (p-UGM) are pairs of the 
format $(G, k)$, where $G$ is an unweighted graph and $k$ is an integer (the parameter).  An algorithm that 
solves the p-UGM problem on an input $(G, k)$, either returns a $k$-matching in the graph $G$, or reports 
that no $k$-matching exists in $G$. 

Throughout this paper, we will let $N = |V| + |E|$ be the ``size'' of a graph $G = (V, E)$.  

We remark that the trivial greedy algorithm that finds a maximal matching, i.e., the algorithm that repeatedly 
adds edges with uncovered ends to the matching, cannot be directly used in our model: to check if an end of an 
edge is uncovered, we need to search in the vertices that are already covered, which will take time upto 
$O(\log k)$, resulting in an algorithm whose running time is at least $O(N \log k)$.  

Let $G$ be a graph and let $k$ be an integer. A vertex $v$ in $G$ is a {\it large-vertex} if the degree 
of $v$ is not smaller than $2k$. A vertex is a {\it small-vertex} if it is not a large-vertex.

\begin{lemma}
\label{lem21}
If a graph $G$ has at least $k$ large-vertices, then $G$ has a $k$-matching, which can be constructed in 
time $O(N + k^2 \log k)$  and space $O(k)$.

\begin{proof}
Let $v_1$, $v_2$, $\ldots$, $v_k$ be $k$ large-vertices in $G$. We simply pick $k$ edges 
of the form $[v_i, w_i]$, where for each $1 \leq i \leq k$, the vertex $w_i$ is not in the vertex set 
$Q = \{v_1, v_2, \ldots, v_k \} \cup \{w_1, w_2, \ldots, w_{i-1} \}$. Note that this is always possible 
since the large-vertex $v_i$ has at least $2k$ neighbors while the number of vertices in the set 
$Q \setminus \{v_i\}$ is $(k-1) + (i-1) \leq 2k - 2$. Such $k$ edges $[v_i, w_i]$, $1 \leq i \leq k$, 
obviously make a $k$-matching in the graph $G$.

To implement this, we scan the graph $G$ to identify the first $k$ large-vertices $v_1$, $v_2$, $\ldots$, 
$v_k$ in $G$ and store them in the set $Q$ in space $O(k)$. The set $Q$ is organized as a balance search 
tree that supports searching and insertion in logorithmic time per operation. We then re-scan the graph 
$G$, and for each $i$, $1 \leq i \leq k$, we work on the large-vertex $v_i$. Inductively, we have the 
set $Q = \{v_1, v_2, \ldots, v_k \} \cup \{w_1, w_2, \ldots, w_{i-1} \}$ stored in space $O(k)$. Since the set 
$Q \setminus \{v_i\}$ has no more than $2k-2$ vertices, to find an edge 
$[v_i, w_i]$ where $w_i$ is not in the vertex set $Q$, we need to examine at most $2k-1$ neighbors of 
$v_i$. After finding the edge $[v_i, w_i]$, we add the vertex $w_i$ to the set $Q$, thus completing the 
process on the $i$-th large-vertex $v_i$. As a result, finding the edge $[v_i, w_i]$ takes at most $O(k)$ 
searching/insertion operations on the set $Q$, which is done in time $O(k \log k)$. In conclusion, it 
takes time $O(N + k^2 \log k)$ and space $O(k)$ to construct the $k$-matching 
$\{ [v_1, w_1], [v_2, w_2], \ldots, [v_k, w_k] \}$ in the graph $G$. 
\end{proof}
\end{lemma}

Now we consider the situation where the graph $G$ has only $h$ large-vertices $v_1$, $v_2$, $\ldots$, $v_h$, 
where $h < k$. An {\it $h$-reduced graph $G_h$} of $G$ is constructed from $G$, using the following procedure: 
\begin{quote}
1. For each large-vertex $v_i$: pick arbitrary $\mbox{deg}(v_i) - 2k$ edges of the form $[v_i, w_i]$, \\
\hspace*{4mm} where $w_i$ is a small-vertex, and delete these edges. \\
2. Delete all vertices of degree $0$ in the resulting graph.
\end{quote}

We give some remarks on the $h$-reduced graph $G_h$. First, for each large-vertex $v_i$, it is always possible 
to find $\mbox{deg}(v_i) - 2k$ edges of the form $[v_i, w_i]$, where $w_i$ is a small-vertex. This is because 
$v_i$ has at least $2k$ neighbors while there are only $h < k$ large-vertices. Secondly, since we only delete edges 
whose one end is a large-vertex and the other end is a small-vertex, when we delete edges incident to a large-vertex, 
no other large-vertices would change their degrees. In particular, all large-vertices in the $h$-reduced graph $G_h$ 
has degree exactly $2k$.  

\begin{lemma}
\label{lem22}
Let $G$ be a graph that has $h$ large-vertices $v_1$, $v_2$, $\ldots$, $v_h$, with $h < k$, and let $G_h$ be 
an $h$-reduced graph of $G$. Then the graph $G$ has a $k$-matching if and only if the $h$-reduced graph 
$G_h$ has a $k$-matching.

\begin{proof}
Since the $h$-reduced graph $G_h$ is a subgraph of the graph $G$, if $G_h$ has a $k$-matching, then 
obviously the graph $G$ has a $k$-matching.

To prove the other direction, assume, to the contrary, that the graph $G$ has a $k$-matching but the 
$h$-reduced graph $G_h$ has no $k$-matching. Suppose that a $k$-matching in $G$ can have at most $r$ edges 
in the $h$-reduced graph $G_h$. Thus, $r < k$. Let ${\cal A}_r$ be the set of all $k$-matchings in $G$ 
that have exactly $r$ edges in the $h$-reduced graph $G_h$. We first study the properties of 
$k$-matchings in the set ${\cal A}_r$. Let $M$ be any $k$-matching in ${\cal A}_r$. Since the 
graph $G_h$ has no $k$-matching, there is at least one edge $e_0$ in $M$ that is not in $G_h$. 

(1) The $k$-matching $M$ must cover all large-vertices. To see this, suppose that $M$ does 
not cover a large-vertex $v_i$. Consider the $(k-1)$-matching $M_0^- = M \setminus \{e_0\}$, 
where $e_0$ is an edge in $M$ that is not in $G_h$. The $(k-1)$-matching $M_0^-$ also contains $r$ 
edges in the $h$-reduced graph $G_h$. There are at most $2k - 2$ neighbors of the large-vertex 
$v_i$ in $G_h$ that are covered by the $(k-1)$-matching $M_0^-$. Since the large-vertex $v_i$ has 
$2k$ neighbors in $G_h$, there is  a neighbor $w_i$ of $v_i$ in $G_h$ that is not covered by $M_0^-$. 
Therefore, $M_0^- \cup \{[v_i, w_i]\}$ gives a $k$-matching in $G$ that has $r+1$ edges in $G_h$, 
contradicting the assumption that a $k$-matching in $G$ can have at most $r$ edges in $G_h$. 
This contradiction proves that the $k$-matching $M$ must cover all large-vertices. 

(2) The $k$-matching $M$ does not contain edges whose both ends are large-vertices. Suppose that 
$M$ contains an edge $e_1 = [v_i, v_j]$ whose both ends $v_i$ and $v_j$ are large-vertices. First note 
that the edge $e_1$ must be in the graph $G_h$ since in the construction of the $h$-reduced graph 
$G_h$, we never delete edges whose both ends are large-vertices. Since at most $2k-2$ neighbors of 
the large-vertex $v_i$ can be covered by the $(k-1)$-matching $M_1^- =M \setminus \{e_1\}$ 
and since the large-vertex $v_i$ has $2k$ neighbors in $G_h$, at least one neighbor $w_i \neq v_j$ 
of $v_i$ in $G_h$ is not covered by $M_1^-$. Thus, replacing the edge $e_1 = [v_i, v_j]$ by the 
edge $[v_i, w_i]$ gives a $k$-matching that has $r$ edges in $G_h$ but leaves the large-vertex 
$v_j$ uncovered. But this contradicts what we have proved in (1) that a $k$-matching in ${\cal A}_r$ 
must cover all large-vertices. 

(3) The $k$-matching $M$ cannot contain an edge $e_2 = [v_i, x_i]$ in $G$ that is not in the $h$-reduced 
graph $G_h$, where $v_i$ is a large-vertex. Again if such an edge $e_2$ exists, then there must be a neighbor 
$w_i$ of $v_i$ in $G_h$ such that $w_i$ is not covered by the $(k-1)$-matching $M_2^- = M \setminus \{e_2\}$. 
Thus, the $k$-matching $M_2^- \cup \{[v_i, w_i]\}$ would give a $k$-matching in $G$ that has $r+1$ edges in 
the $h$-reduced graph $G_h$, contradicting the definition of $r$.  

Summarizing (1)-(3), we conclude that the $k$-matching $M$ must contain $h$ edges in the $h$-reduced 
graph $G_h$, with one end being a large-vertex and the other end being a small-vertex. Since there are 
only $h$ large-vertices in the graph $G$, the other $k-h$ edges in $M$ must have their both ends being 
small-vertices. Because in the construction of the $h$-reduced graph $G_h$, we never delete edges 
whose both ends are small-vertices, these $k-h$ edges in $M$ must also be in the $h$-reduced graph 
$G_h$. Thus, the $k$-matching $M$ in $G$ is a $k$-matching in the $h$-reduced graph $G_h$, 
contradicting the assumption that $G_h$ has no $k$-matching, thus, proving the lemma.  
\end{proof}
\end{lemma}

By Lemma~\ref{lem22}, it suffices to consider how to construct a $k$-matching in the $h$-reduced graph $G_h$. 
Unfortunately, because of the space limit, we cannot construct the $h$-reduced graph explicitly. In the following, 
we show how we can construct a $k$-matching in an ``implicit'' $h$-reduced graph $G_h$. For simplicity, we will 
call an edge $e$ a {\it small-edge} if both ends of $e$ are small-vertices. 

\begin{lemma}
\label{lem23}
Let $h < k$. If the $h$-reduced subgraph $G_h$ has a subgraph $G_h'$ that contains all edges that are incident to 
the large-vertices in $G_h$ and at least $(4k-3)(k-h)$ small-edges in $G_h$, then $G_h'$ has a $k$-matching that 
can be constructed in time $O(k^2 \log k)$ and space $O(k^2)$. 

\begin{proof}
First note that the graph $G_h'$ can be stored in space $O(k^2)$. We construct a $k$-matching in the graph 
$G_h'$, as follows: (1) start with an empty matching $M$; and (2) repeatedly pick an edge $e$ from the remaining 
small-edges, include $e$ in the matching $M$, and delete the two ends of $e$ (and all incident edges). Since 
there are at most $4k-4$ other small-edges that can share common ends with $e$, with the $(4k-3)(k-h)$ 
small-edges in $G_h'$, we will be able to construct a matching of $k-h$ edges in $G_h'$. Now, as we did in 
Lemma~\ref{lem21}, we proceed with each $v_i$ of the $h$ large-vertices $\{v_1, \ldots, v_h\}$, where we can 
find an edge $[v_i, w_i]$, where $w_i$ is a small-vertex not covered by $M$, so we can add the edge $[v_i, w_i]$ 
to the matching $M$. This gives a $k$-matching $M$ in the graph $G_h'$. 

To achieve the time complexity given in the lemma, we store the edges and vertices of the graph $G_h'$ in 
balanced search trees so that searching, insertion, and deletion take $O(\log k)$ time per operation, which leads 
to the $O(k^2 \log k)$ running time of the algorithm. 
\end{proof}
\end{lemma}

Now we are ready for our matching algorithm for unweighted graphs, as given in Figure~\ref{alg1}, where  
Best-Match in step 5 is an algorithm that solves the $k$-matching problem in the $h$-reduced subgraph $G_h$, 
whose complexity will be discussed in detail later. In order to keep the running time of the algorithm {\bf UGM} 
to be linear in terms of the input size $N$, we need to use certain randomness, which will be explained in 
the proof of Theorem~\ref{theo1}. Thus, our algorithm is a randomized algorithm, whose error bound 
and complexity are given in the following theorem.  
\begin{figure}[htbp]
\setbox4=\vbox{\hsize30pc
\medskip
 \noindent\strut  
\hspace*{3mm}\footnotesize {\bf Algorithm UGM}\\
\hspace*{3mm}{\sc input}:  an unweighted graph $G$ and parameter $k$\\
\hspace*{3mm}{\sc output}: a $k$-matching in $G$, or report no such a matching in $G$. 

\smallskip
\hspace*{3mm}1. \hspace*{1mm} collect upto $k$ large-vertices in $G$, store them in $V_L$; let $h = |V_L|$;\\
\hspace*{3mm}2. \hspace*{1mm} {\bf if} $(h = k)$ {\bf return} a $k$-matching $M$ in $G$;\\
\hspace*{3mm}3. \hspace*{1mm} construct the $h$-reduced graph $G_h$ but keep upto $(4k-3)(k-h)$ small-edges; \\
\hspace*{3mm}4. \hspace*{1mm} {\bf if} ($G_h$ has $(4k-3)(k-h)$ small-edges) {\bf return} a 
        $k$-matching $M$ in $G$;\\
\hspace*{3mm}5. \hspace*{1mm} call Best-Match$(G_h)$ to solve the problem. 
\medskip
\strut} $$\boxit{\box4}$$
 \vspace{-8mm}
\caption{The $k$-matching algorithm for unweighted graphs} 
\label{alg1}
\end{figure}

\begin{theorem}
\label{theo1}
For any $\epsilon > 0$, with probability at least $1 - \epsilon$, the algorithm {\bf UGM} solves the p-UGM problem 
in time $O(N + k^2 \log k + k \log(1/\epsilon) + \alpha(k^2))$ and space $O(k^2)$, where $\alpha(k^2)$ is the time 
complexity for finding a $k$-matching in a graph of $O(k^2)$ edges and without degree-$0$ vertices, with the space 
complexity simultaneously bounded by $O(k^2)$.

\begin{proof}
The correctness of the algorithm is obvious: Lemma~\ref{lem21} and Lemma~\ref{lem23} ensure, respectively, that 
if the algorithm returns at step 2 and step 4, then it returns a $k$-matching in the graph $G$. If the algorithm returns 
from step 5, then Lemma~\ref{lem22} guarantees that the algorithm returns a $k$-matching in the $h$-reduced graph 
$G_h$, which is also a $k$-matching in the original graph $G$, if and only if the original graph $G$ has $k$-matchings. 

We study the complexity of the algorithm. Recall that the graph $G$ is given in an adjacency list. Thus, the degree 
of a vertex can be computed by reading the list of neighbors of the vertex. As a result, step 1 takes time $O(N)$. Since 
we keep at most $k$ large-vertices of $G$ in the set $V_L$, the set $V_L$ can be stored in space $O(k)$. In case the 
number $h$ of large-vertices in the set $V_L$ is equal to $k$, by Lemma~\ref{lem21}, step 2 of the algorithm constructs 
a $k$-matching $M$ in $G$ in time $O(N + k^2 \log k)$ and space $O(k)$, and returns. 

If the number $h$ of large-vertices in the set $V_L$ is smaller than $k$, then step 3 of the algorithm constructs the 
$h$-reduced graph $G_h$. For this, we need to be more careful: in order to collect the small-edges, we need to decide 
for each edge if any end of the edge is a large-vertex. Even if we organize the large-vertices in a balanced search tree, 
it would still take time $O(N \log h) = O(N \log k)$ to go through the edges of $G$ and construct the $h$-reduced graph 
$G_h$. 

To solve this problem, we use the technique of unversal hashing. For the set $V_L$ of the $h$ collected large-vertices, 
we  pick a hash function $H$ from $U$ to $[1 .. h^2]$ randomly from a universal class of hash functions, where $U$ 
is the set of the vertices in the input graph $G$. With a probabililty at least $1/2$, the function $H$ is injective from the 
set $V_L$ to $[1 .. h^2]$ (see \cite{clrs}, Theorem 11.9).  The hash function $H$ can be constructed in constant  
(randomized) time. Moreover, after initializing an array $A[1 .. h^2]$ in time $O(h^2)$, we can check if the function 
$H$ is injective from $V_L$ in time $O(h)$: for this, we fix a distinct value $a_H$ for the function $H$, and 
for each vertex $v$ in $V_L$, if $A[H[v]]$ is already equal to $a_H$, then the function $H$ is not injective from $V_L$, 
otherwise, we set $A[H[v]] = a_H$. Therefore, for any $\epsilon > 0$, by repeating this procedure $\log(1/\epsilon)$ 
times, thus in time $O(\log(1/\epsilon)h + h^2)$ (note that the array $A[1 .. h^2]$ needs to be initialized only once), 
with probability at least $1 - \epsilon$, we will get a hash function $H_0$ that is injective from $V_L$. Using this hash 
function $H_0$, we re-initialize the array $A[1 .. h^2]$, and then place the $h$ large-vertices in $V_L$ in the array 
$A[1 .. h^2]$ so that a large-vertex $w$ is placed in $A[H_0[w]]$. Since $h < k$, we conclude that with probability 
at least $1 - \epsilon$ and in time $O(\log(1/\epsilon)k + k^2)$ and space $O(k^2)$ (which is mainly for the array 
$A[1 .. h^2]$), we will find the hash function $H_0$ that is injective from $V_L$ and finalize the array $A[1 .. h^2]$. 
Now for any vertex $v$ in the input graph $G$, by checking the value $A[H_0(v)]$, which takes constant time, we 
can easily find out if $v$ is a large-vertex. 
 
Now it is straightforward to construct the $h$-reduced graph $G_h$. We simply scan the input graph $G$. For each 
large-vertex $v_i$, we delete all but $2k$ edges incident to $v_i$, (keeping all the edges of the form $[v_i, w]$ where 
$w$ is a large-vertex), and for each small-vertex 
$w$, we record the small-edges incident to $w$. The process stops either when we have collected $(4k-3)(k-h)$ 
small-edges, or when all edges of the graph $G$ are examined. In the former case, we get a subgraph $G_h'$ of 
the $h$-reduced graph $G_h$ that satisfies the conditions of Lemma~\ref{lem23}, thus, step 4 of the algorithm 
{\bf UGM} constructs a $k$-matching $M$ in $G_h'$ (thus also in $G_h$ and in $G$) in time $O(k^2 \log k)$ and 
space $O(k^2)$. In the latter case, the $h$-reduced graph $G_h$ has fewer than $2kh + (4k-3)(k-h) = O(k^2)$ 
edges, so the algorithm Best-Match in step 5 is applied on the graph $G_h$ with $O(k^2)$ edges and Lemma~\ref{lem22} 
guarantees the correctness of the algorithm {\bf UGM}.  

We remark that in this process, the vertices in $G$ that become of degree-$0$ after the construction of the 
$h$-reduced subgraph $G_h$ can also be efficiently identified and deleted: for each small-vertex $w$, we do not 
record {\it any} of its incident edges whose other end is a large-vertex. In particular, small-vertices in $G$ that 
are adjacent to only large-vertices are not recorded in this scanning phase. Only after this scanning phase, we 
re-examine the chosen edges incident to large-vertices in the $h$-reduced graph $G_h$, and add further 
small-vertices to $G_h$ if they are the other ends of these edges and are not recorded in the scanning phase. 
This prevents the graph $G_h$ from having degree-$0$ vertices. Thus, the $h$-reduced graph $G_h$ in step 5 has 
$O(k^2)$ edges and has no vertices of degree $0$. As a result, the number $n_h$ of vertices in the $h$-reduced 
subgraph $G_h$ is also bounded by $O(k^2)$. Now we rename the vertices of $G_h$ as integers in $[1 .. n_h]$ 
so that the Best-Match algorithm in step 5 can be applied. This takes another $O(k^2 \log k)$ time and space 
$O(k^2)$. By the assumption, the p-UGM problem on the graph $G_h$ (thus by Lemma~\ref{lem22} 
on the input graph $G$) can be solved in time $\alpha(k^2)$ and space $O(k^2)$. 

Summarizing all the above discussions completes the proof of the lemma. 
\end{proof}
\end{theorem}

Now we study the time complexity $\alpha(k^2)$ of solving the $k$-matching problem in a graph with $O(k^2)$ edges 
(we will assume, without loss of generality, that graphs have no degree-$0$ vertices). There has been extensive 
research on algorithms for constructing a maximum matching in an unweighted graph \cite{hk,bestmatch,bestmatch1}. 
In particular, it is known \cite{bestmatch} that for a graph of $n$ vertices and $m$ edges, a maximum matching in 
the graph can be constructed in time $O(m \sqrt{n})$, from which the $k$-matching problem can be solved trivially. 
Therefore, for graphs of $O(k^2)$ edges, which may have up to $O(k^2)$ vertices, the $k$-matching problem can 
be solved in time $O(k^3)$, giving an upper bound $O(k^3)$ for the complexity $\alpha(k^2)$. In the following, 
we show how a better upper bound for the time complexity $\alpha(k^2)$ can be obtained.

Let $M$ be a matching in a graph $G$. An {\it augmenting path} $P$ (relative to $M$) in $G$ is a simple path whose 
both ends are uncovered by $M$, and whose edges are alternatively going between not in $M$ and in $M$. An 
augmenting path is the {\it shortest} if its length is the minimum over all augmenting paths relative to $M$.  

We start with the following theorem, which is also of its independent interests.

\begin{theorem}
\label{theo2}
There is an $O(m \sqrt{k})$-time and $O(m)$-space algorithm that on a graph $G$ of $m$ edges, either 
constructs a $k$-matching in $G$ or reports that no $k$-matching exists in $G$. 

\begin{proof}
We first prove the following claim:

\medskip

\noindent {\bf Claim.} \ Let $G$ be a graph of $m$ edges, and let $k_0$ be the size of a maximum matching in 
$G$. A maximum matching in the graph $G$ can be constructed in time $O(m \sqrt{k_0})$ and space $O(m)$.

\smallskip

\noindent {\it Proof of the Claim.} 
An algorithm proposed by Micali and Vazirani \cite{bestmatch} constructs a maximum matching in a general graph 
$G$ of $n$ vertices and $m$ edges in time $O(m \sqrt{n})$ and space $O(m)$ (we will call this algorithm the 
{\it MV-algorithm}). The MV-algorithm runs in phases. Each phase starts with a matching $M$, finds a maximal set 
of vertex-disjoint shortest augmenting paths relative to $M$, and augments along all these paths to get a larger 
matching. As proved by Hopcroft and Karp (Theorem 3 in \cite{hk}), running the MV-algorithm for at most 
$2\sqrt{k_0} + 1$ such phases will be sufficient to find a maximum matching in the (general) graph $G$. Moreover, 
Micali and Vazirani \cite{bestmatch} presented an $O(m)$-time algorithm (thus also in space $O(m)$) that 
implements the process of each phase in the MV-algorithm \footnote{This $O(m)$-time algorithm for each phase in 
the MV-algorithm is highly nontrivial. For much more details and discussions, see \cite{bestmatch1,bestmatch2}. 
On the other hand, for bipartite graphs, there is a much simpler $O(m)$-time algorithm that implements the process 
of each phase. See \cite{hk}.}. Combining these two results, we obtain an algorithm that finds a maximum matching 
in a general graph $G$ of $m$ edges in time $O(m \sqrt{k_0})$ and space $O(m)$. This proves the claim.

\medskip

Let us now get back to the proof of the original theorem. Our algorithm proceeds as follows. We first use a trivial 
greedy algorithm to construct a maximal matching $M'$ in the graph $G$ in time $O(m)$ and space $O(m)$. If 
$|M'| \geq k$, then we can easily have a $k$-matching of $G$ from $M'$. On the other hand, we have $|M'| < k$. 
It is well-known that for a graph the size of a maximum matching is at most twice of that of a maximal matching 
\cite{mmvc}. Therefore, if $|M'| < k$, then the maximum matching in the graph $G$ has its size $k_0$ bounded 
by $2k$, and we can apply the above claim to construct a maximum matching $M''$ in $G$ in time 
$O(m \sqrt{2k}) = O(m \sqrt{k})$ and space $O(m)$. Now from the maximum matching $M''$, we can easily 
either construct a $k$-matching in $G$ or report that the graph $G$ has no $k$-matching. This proves the theorem.
\end{proof}
\end{theorem}

By Theorem~\ref{theo2}, we get an upper bound $O(k^{2.5})$ on the time complexity $\alpha(k^2)$ given in 
Theorem~\ref{theo1} for the algorithm {\bf UGM}. Now if we replace $\alpha(k^2)$ with $k^{2.5}$, and let 
$\epsilon = 1/2^{k^{1.5}}$, then Theorem~\ref{theo1} reads as 

\begin{theorem}
\label{theo4}
With probability at least $1 - 1/2^{k^{1.5}}$, the algorithm {\bf UGM} solves the p-UGM problem on general 
graphs in time $O(N + k^{2.5})$ and space $O(k^2)$.
\end{theorem}

Note that the bound $O(N + k^{2.5})$ in Theorem~\ref{theo4} is the best possible for the p-UGM problem 
based on the current status of the research on graph matching algorithms -- the best known algorithm for the graph 
matching problem runs in time $O(n^{2.5})$ on a graph of $n$ vertices \cite{bestmatch}.\footnote{We remark 
that there is a randomized algorithm of time $O(n^{2.376})$ for the graph matching problem, based on fast matrix 
multiplication algorithms \cite{mucha}. However, our $h$-reduced subgraph $G_h$ may have up to $\Omega(k^2)$ 
vertices. Therefore, a direct application of the algorithm in \cite{mucha} would not lead to a faster algorithm for the 
p-UGM problem. Moreover, using the algorithm in \cite{mucha} would require space $O(k^4)$.} 

We are not aware of any parameterized algorithms published in the literature that are specifically for solving the p-UGM 
problem. On the other hand, in the research on streaming algorithms, the p-UGM problem has been studied recently. 
In particular, Chitnis {\it et al.}~\cite{soda2016} presented two randomized algorithms for the p-UGM problem on the 
dynamic graph streaming model. In order to deal with edge deletions in streaming, the algorithms given in \cite{soda2016} 
smartly employed powerful techniques in $l_0$-sampling \cite{l0sampling}. However, these techniques are relatively 
expensive. If we remove these expensive operations, the algorithms givein in \cite{soda2016} can be used to solve 
the p-UGM problem (in the insert-only graph streaming model). With the simplifications, the first algorithm given in 
\cite{soda2016}, for any $\epsilon > 0$, runs in time $O(N \log(1/\epsilon) + \beta(k))$ and converts an input graph 
$G$ of size $N$ into a graph of up to $O(k^4 \log(1/\epsilon))$ edges with a probability $1 - \epsilon$. Thus, both 
the bound $\beta(k)$ in the time complexity and the space complexity of the algorithm are at least 
$O(k^4 \log(1/\epsilon))$. Moreover, to achieve a probability $1 - o(1)$, the algorithm would require super-linear time. 
The second algorithm given in \cite{soda2016}, if simplified as described above, converts a graph $G$ of size $N$ into 
a graph with $O(k^2 \log(1/\epsilon))$ edges. The algorithm runs in time $O(N \log k)$ even if we only want to achieve 
a probability $1 - \epsilon$ for a constant $\epsilon > 0$. More seriously, the algorithm only applies to graphs 
in which the size of a maximum matching is bounded by $O(k)$.

\section{Case study II: matching in weighted graphs}

In this section, we study the maximum weighted $k$-matching problem on weighted graphs, i.e., the {\it p-WGM 
problem}. Let $G$ be a weighted graph. A {\it maximum $k$-matching} in $G$ is a $k$-matching in $G$ whose 
weight is the largest over all $k$-matchings in $G$. The instances of the p-WGM problem consist of pairs of the 
form $(G, k)$, where $G$ is a weighted graph and $k$ is an integer. A solution to the instance $(G, k)$ is either 
a maximum $k$-matching in $G$ or a report that no $k$-matching exists in $G$.

We remark that in practice, the p-WGM problem is probably applicable to more applications, compared to the 
p-UGM problem. Indeed, with a very large graph $G$, we may only be interested in having a certain number $k$ 
of matched vertex pairs where $k$ is not necessarily the largest. On the other hand, we may want to have $k$ 
such matched pairs that maximize an objective value.  

Technically, the p-WGM problem becomes very different from the p-UGM problem. A weighted graph $G$ may 
have matchings of very large size (i.e., the number of edges in the matching) while we are just looking for a 
maximum $k$-matching where $k$ could be relatively small. In particular, Lemmas~\ref{lem21}-\ref{lem23} are 
no longer useful because now the graph $G$ may have a very large number of large-vertices, and, even for a 
subgraph with a very large number of edges, there is no guarantee that the subgraph would contain a maximum 
$k$-matching in the original graph. Finally, the technique we used in the proof of Theorem~\ref{theo1} to pre-scan 
the graph $G$ and collect the large-vertices cannot be used -- there can be simply too many large-vertices. 

We start with the following lemma that will be useful in several places in our construction.

\begin{lemma}
\label{lem31}
There is an algorithm that on an input of $n$ elements and a parameter $k$, produces the $k$ largest elements 
in the input in time $O(n)$ and space $O(k)$.

\begin{proof}
The algorithm starts by reading the first $k$ elements $\{ a_1, a_2, \ldots, a_k \}$ from the input. Inductively, suppose 
that for an integer $i \geq k$, the algorithm has obtained the $k$ largest elements $b_1$, $b_2$, $\ldots$, $b_k$ in the 
first $i$ elements in the input. The algorithm then reads the next block $\{ a_{i+1}, a_{i+2}, \ldots, a_{i+k} \}$ of $k$ 
elements in the input, and use the linear-time Median-Finding algorithm \cite{clrs} to find the $k$-th largest element in the 
set $S_{i+k} = \{b_1, b_2, \ldots, b_k, a_{i+1}, a_{i+2}, \ldots, a_{i+k}\}$ in time $O(k)$, from which the $k$ largest 
elements in the set $S_{i+k}$, which are also the $k$ largest elements in the first $i+k$ elements in the input, can be 
easily obtained. Since the algorithm spends time $O(k)$ on each block of $k$ elements in the input, we conclude that 
the running time of the algorithm is $O(n)$. Moreover, it is obvious that the algorithm takes $O(k)$ space. 
\end{proof}
\end{lemma}

Let $G$ be a weighted graph. Similarly (but not identically) to the process on the problem p-UGM, we define a 
{\it large-vertex} to be a vertex whose degree is at least $8k$ and a {\it small-vertex} to be a vertex whose 
degree is less than $8k$. In the following, we will introduce operations that remove edges from the weighted 
graph $G$ without changing the weight of its maximum $k$-matchings. This will require the condition that each 
edge in the weighted graph $G$ have a distinct weight, which, in general, is not the case. For this, we introduce 
a new edge weight function for the graph $G$ as follows: let $e = [v, w]$ be an edge of weight $wt(e)$ in the graph 
$G$, we define the new weight $wt'(e)$ for the edge $e$ as a triple $wt'(e) = (wt(e), \min\{v, w\}, \max\{v, w\})$. 
The new edge weights follow the lexicographic order. In terms of the weight function $wt'(\;)$, each edge in the 
graph $G$ has a distinct weight. Moreover, for any edge set $S$ and any integer $h$, the set of the $h$ heaviest 
edges in $S$ in terms of the weight function $wt'(\;)$, which is uniquely defined, must be a set that consists of $h$ 
heaviest edges in $S$ in terms of the weight function $wt(\;)$. 

We first consider the following two kinds of subgraphs constructed from the weighted graph $G$, where the edge 
weights are measured by the new edge weights $wt'(\cdot)$ as defined above: 
\begin{itemize} 
   \item The {\it trimmed subgraph $G_T$} of the graph $G$ consists of the edges  
      $e = [v, w]$ in $G$ such that $e$ is among the $8k$ heaviest edges incident to the vertex $v$ and among the 
      $8k$ heaviest edges incident to the vertex $w$, plus the vertices incident to these edges.  

\vspace*{-2mm} 

     \item The {\it reduced subgraph $G_R$} of $G$ is a subgraph of the trimmed subgraph $G_T$ of $G$ such that 
         either $G_R = G_T$ if $G_T$ has no more than $k(16k-1)$ edges, or $G_R$ consists of the $k (16k-1)$ heaviest 
         edges in $G_T$, plus the vertices incident to the edges. 
\end{itemize}

{\bf Remark 1.} Note that every edge incident to a small-vertex $v$ is among the $8k$ heaviest edges incident to 
the vertex $v$. 

{\bf Remark 2.} Because each edge $e$ in the graph $G$ has a distinct edge weight $wt'(e)$, the trimmed subgraph 
$G_T$ and the reduced subgraph $G_R$ of the graph $G$ are uniquely defined. 

{\bf Remark 3.} Each vertex in the trimmed subgraph $G_T$, thus also each vertex in the reduced subgraph $G_R$, 
has degree bounded by $8k$. Note that a large-vertex $v$ in the graph $G$ may have degree less than $8k$ in the 
trimmed subgraph $G_T$. In particular, if an edge $e = [v, w]$ is among the $8k$ heaviest edges incident to $v$ 
but not among the $8k$ heaviest edges incident to $w$, then the degree of the vertex $v$ in the trimmed subgraph 
$G_T$ is less than $8k$. 

{\bf Remark 4.} The size of the trimmed subgraph $G_T$ can still be very large (since there can be many 
large-vertices). On the other hand, the reduced subgraph $G_R$ has size bounded by $O(k^2)$. 

\begin{lemma}
\label{lem321}
A maximum $k$-matching in the trimmed subgraph $G_T$ of a weighted graph $G$ is also a maximum $k$-matching 
in the original graph $G$.\footnote{Note that although when we compare edges we use the new weight function 
$wt'(\;)$, the weight of a matching is still defined in terms of the original edge weight function $wt(\;)$.} 

\begin{proof}
For each large-vertex $v$ in the graph $G$, let $e_{8k}(v)$ be the $(8k)$-th heaviest edge incident to $v$. Consider 
the algorithm in Figure~\ref{alg321} that constructs the trimmed subgraph $G_T$. 
\begin{figure}[htbp]
\setbox4=\vbox{\hsize30pc
\smallskip
 \noindent\strut  
\hspace*{2mm}\footnotesize {\bf Algorithm Triming}

\hspace*{2mm}1. sort the large-vertices in the weighted graph $G$ in a sequence: $v_1', v_2', \ldots, v_h'$, such \\
\hspace*{5mm} that $wt'(e_{8k}(v_1')) \leq wt'(e_{8k}(v_2')) \leq \cdots \leq wt'(e_{8k}(v_h'))$;\\
\hspace*{2mm}2. {\bf for} $i = 1$ {\bf to} $h$ {\bf do} 
delete all but the $8k$ heaviest edges incident to $v_i'$. 
\smallskip
\strut} $$\boxit{\box4}$$
 \vspace{-8mm}
\caption{Constructing the trimmed subgraph $G_T$ of a weighted graph $G$} 
\label{alg321}
\end{figure}

Since every edge $e$ in the graph $G$ has a distinct edge weight $wt'(e)$, when we delete edges incident to a 
large-vertex $v_i'$, we would not delete any of the $8k$ heaviest edges incident to a large-vertex $v_j'$ with 
$i < j$. Therefore, if we let $G_i$ be the graph $G$ after deleting all but the $8k$ heaviest edges incident to the 
vertex $v_s'$ for all $s \leq i$, then the graph $G_{i+1}$ will be obtained from the graph $G_i$ by deleting all but 
the $8k$ heaviest edges incident to the vertex $v_{i+1}'$, and the graph $G_h$ constructed by the algorithm 
is the trimmed subgraph $G_T$. We prove by induction on $i$ that for all $i$, a maximum $k$-matching in the 
graph $G_i$ is also a maximum $k$-matching in the original graph $G$. This is certainly true for $i = 0$. 

Let $M_i$ be a maximum $k$-matching in the graph $G_i$. Consider the graph $G_{i+1}$ that is obtained from 
$G_i$ by deleting all but the $8k$ heaviest edges incident to the vertex $v_{i+1}'$. If $M_i$ contains no edge that 
is deleted in the construction of $G_{i+1}$ from $G_i$, then $M_i$ is also a matching in $G_{i+1}$. Otherwise, $M_i$ 
contains an edge $[v_{i+1}', w]$ that is not among the $8k$ heaviest edges $[v_{i+1}', w_s]$, $1 \leq s \leq 8k$, 
incident to the vertex $v_{i+1}'$ in the graph $G_i$. Since the $(k-1)$-matching $M_i \setminus \{[v_{i+1}', w]\}$ 
can cover at most $2k-2$ neighbors of $v_{i+1}'$, there must be an edge $[v_{i+1}', w_t]$ among the $8k$ 
heaviest edges incident to $v_{i+1}'$ such that the vertex $w_t$ is not covered by $M_i \setminus \{[v_{i+1}', w]\}$. 
Thus, replacing the edge $[v_{i+1}', w]$ with the edge $[v_{i+1}', w_t]$ will give a $k$-matching $M_i'$ in the graph 
$G_{i+1}$. By the definition of the weight $wt'(\cdot)$, we must have $wt([v_{i+1}', w]) \leq wt([v_{i+1}', w_t])$. 
Thus, the $k$-matching $M_i'$ has a weight at least as large as that of $M_i$. Therefore, the graph $G_{i+1}$ 
always has a $k$-matching whose weight is at least as large as that of the maximum $k$-matching $M_i$ in the 
graph $G_i$. Since $G_{i+1}$ is a subgraph of $G_i$, we conclude that a maximum $k$-matching in the graph 
$G_{i+1}$ is also a maximum $k$-matching in the graph $G_i$, which, by induction, is also a maximum $k$-matching 
in the original graph $G$. 
\end{proof}
\end{lemma}

\begin{lemma}
\label{lem32}
A maximum $k$-matching in the reduced subgraph $G_R$ of a weighted graph $G$ is also a maximum $k$-matching 
in the original graph $G$.

\begin{proof}
By Lemma~\ref{lem321}, it suffices to prove that a maximum $k$-matching $M_R$ in the reduced subgraph $G_R$ 
is also a maximum $k$-matching in the trimmed subgraph $G_T$. If the trimmed subgraph $G_T$ has fewer than 
$k(16k-1)$ edges, then by definition, $G_R = G_T$, and $M_R$ is obviously a maximum $k$-matching in $G_T$. Thus, 
we can assume that the reduced subgraph $G_R$ has exactly $k(16k-1)$ edges, which are the $k(16k-1)$ heaviest 
edges in the trimmed subgraph $G_T$. Let $M_T$ be a maximum $k$-matching in the trimmed subgraph $G_T$. 
Assume that $M_T = M_T' \cup M_T''$, where $M_T'$ is the set of edges that are in the reduced subgraph $G_R$ 
and $M_T''$ is the set of edges that are not in the reduced subgraph $G_R$, with $|M_T'| = h$ and $|M_T''| = k-h > 0$. 
Now for each edge $e$ in $M_T'$, delete the two ends of $e$ (and all incident edges) in the graph $G_R$. Since the 
graph $G_R$ has $k(16k-1)$ edges, and the vertex degree of $G_R$ is bounded by $8k$, this will delete at most 
$h(16k-1)$ edges in $G_R$. Thus, the resulting graph $G_R'$ still has at least $k(16k-1) - h(16k-1) = (k-h)(16k-1)$ edges. 
Now in the graph $G_R'$, because the vertex degree is bounded by $8k$, we can easily construct a $(k-h)$-matching 
$M_R''$ in $G_R'$ (thus in $G_R$) by repeatedly including an (arbitrary) edge in the matching and removing all edges 
incident to the ends of the edge. Since no edge in $M_T''$ is in $G_R$, by the definition of the reduced subgraph 
$G_R$, the weight of the $(k-h)$-matching $M_R''$ in $G_R$ is at least as large as that of the $(k-h)$-matching $M_T''$ 
in $G_T$. Therefore, replacing the $(k-h)$-matching $M_T''$ in $M_T$ with the $(k-h)$-matching $M_R''$ gives a 
$k$-matching $M_T' \cup M_R''$ in the reduced subgraph $G_R$ whose weight is at least as large as that of the maximum 
$k$-matching $M_T$ in the trimmed subgraph $G_T$. As a consequence, the weight of the maximum $k$-matching in 
the reduced subgraph $G_R$ is at least as large as that of the maximum $k$-matching $M_T$ in the trimmed subgraph 
$G_T$. Since $G_R$ is a subgraph of $G_T$, we conclude that a maximum $k$-matching in the reduced subgraph 
$G_R$ is also a maximum $k$-matching in the trimmed subgraph $G_T$.
\end{proof}
\end{lemma}

By Lemma~\ref{lem32}, to construct a maximum $k$-matching in the input graph $G$, it suffices to construct a 
maximum $k$-matching in the reduced subgraph $G_R$, which is a subgraph of the trimmed subgraph $G_T$ and 
has a size $O(k^2)$. However, it seems challenging to construct the reduced subgraph $G_R$ from the weighted 
graph $G$ in time $O(N + k^{O(1)})$ and space $k^{O(1)}$: 
\begin{quote}
(1) The trimmed subgraph $G_T$ can be very large, and we may not have enough space 
\hspace*{5mm} to store the entire trimmed subgraph $G_T$; 

(2) The number of large-vertices can be very large. Although any proper subset of \\
\hspace*{5mm} at least $k$ large-vertices and their incident edges contain a $k$-matching in $G$, there\\
\hspace*{5mm} is no guarantee that the $k$-matching is of the maximum weight. On the other\\
\hspace*{5mm} hand, we may not have enough space to record  all large-vertices

(3) Because of (2), even constructing the trimmed subgraph $G_T$ ``locally'' becomes\\
\hspace*{5mm} difficult: to determine if an edge $e = [v, w]$ of $G$ is in $G_T$, we need to know if 
\hspace*{5mm} $wt'(e) \geq wt'(e_{8k}(v))$ and $wt'(e) \geq wt'(e_{8k}(w))$. Note that this should be done in \\
\hspace*{5mm} constant time in average, in order to achieve the $O(N + k^{O(1)})$ time complexity\\
\hspace*{5mm} for the construction of the reduced subgraph $G_R$;

(4) In order to keep the size of the reduced subgraph $G_R$ by $O(k^2)$, we also need to\\
\hspace*{5mm} exclude the vertices of $G$ that are incident to no edges in $G_R$.  
\end{quote}

We develop new techniques to deal with these technical difficulties. Again for a large-vertex $v$, we let $e_{8k}(v)$ 
be the $(8k)$-th heaviest edge incident to $v$ in the graph $G$, in terms of the weight function $wt'(\;)$. The value 
$wt'(e_{8k}(v))$ will be called the {\it $e_{8k}$-value} of the vertex $v$. For the convenience of discussions, we 
define the $e_{8k}$-value of a small-vertex to be $-\infty$.   

The {\it bounding set $B_{8k}$} of large-vertices in the graph $G$ is defined as follows:

\smallskip

(1) if there are at most $8k$ large-vertices in $G$, then $B_{8k}$ contains all large-vertices; and

(2) if there are more than $8k$ large-vertices in $G$, then $B_{8k}$ contains the $8k$ large-vertices\\ 
\hspace*{11mm} whose $e_{8k}$-values are among the $8k$ largest $e_{8k}$-values over all large-vertices of $G$. 

\smallskip

\noindent Similarly, for a vertex $v$, we define the {\it bounding list $L_v^{8k}$} of edges incident to $v$ as following: 

\smallskip

(1) if $v$ is a small-vertex, then $L_v^{8k}$ consists of all edges incident to $v$; and 

(2) if $v$ is a large vertex, then $L_v^{8k}$ consists of the $8k$ heaviest edges incident to $v$. 

\smallskip

\begin{figure}[htbp]
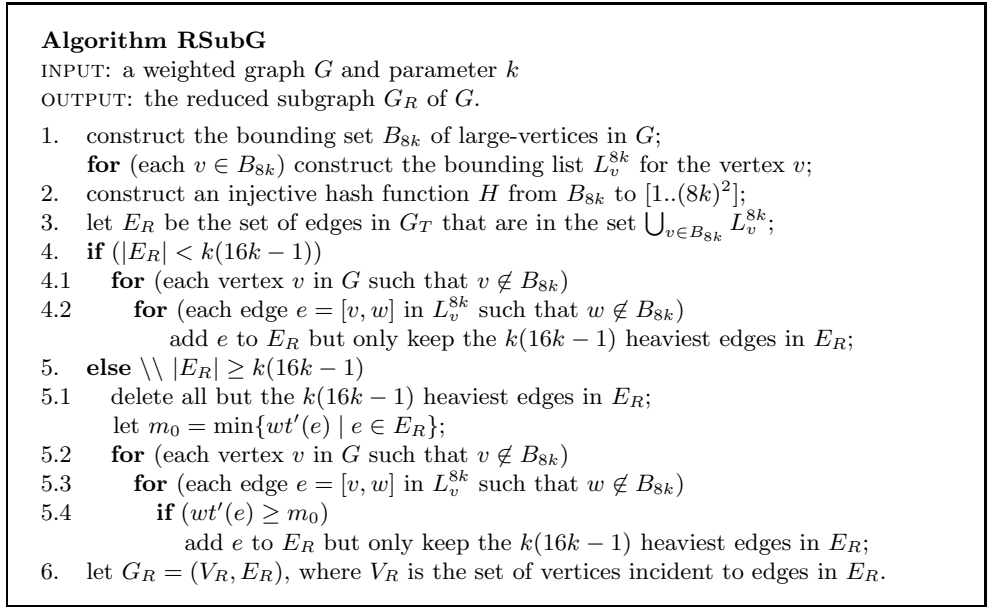

\setbox4=\vbox{\hsize30pc
\medskip
 \noindent\strut  
\hspace*{3mm}\footnotesize {\bf Algorithm RSubG}\\
\hspace*{3mm}{\sc input}:  a weighted graph $G$ and parameter $k$\\
\hspace*{3mm}{\sc output}: the reduced subgraph $G_R$ of $G$. 

\smallskip
\hspace*{3mm}1. \hspace*{1mm}  construct the bounding set $B_{8k}$ of large-vertices in $G$;\\
\hspace*{8mm} {\bf for} (each $v \in B_{8k}$) construct the bounding list $L_v^{8k}$ for the vertex $v$; \\
\hspace*{3mm}2. \hspace*{1mm}  construct an injective hash function $H$ from $B_{8k}$ to $[1 .. (8k)^2]$;\\
\hspace*{3mm}3. \hspace*{1mm} let $E_R$ be the set of edges in $G_T$ that are in the set 
             $\bigcup_{v \in B_{8k}} L_v^{8k}$;\\ 
\hspace*{3mm}4. \hspace*{1mm} {\bf if} ($|E_R| < k(16k - 1)$) \\
\hspace*{3mm}4.1 \hspace*{3mm} {\bf for} (each vertex $v$ in $G$ such that $v \not\in B_{8k}$) \\
\hspace*{3mm}4.2 \hspace*{6mm} {\bf for} (each edge $e = [v, w]$ in $L_v^{8k}$ such that $w \not\in B_{8k}$)\\
\hspace*{19mm} add $e$ to $E_R$ but only keep the $k(16k-1)$ heaviest edges in $E_R$;\\
\hspace*{3mm}5. \hspace*{1mm} {\bf else} $\backslash\backslash$ $|E_R| \geq k(16k - 1)$\\
\hspace*{3mm}5.1 \hspace*{3mm} delete all but the $k(16k-1)$ heaviest edges in $E_R$; \\
\hspace*{11.5mm} let $m_0 = \min\{wt'(e) \mid e \in E_R\}$; \\
\hspace*{3mm}5.2 \hspace*{3mm} {\bf for} (each vertex $v$ in $G$ such that $v \not\in B_{8k}$) \\
\hspace*{3mm}5.3 \hspace*{6mm} {\bf for} (each edge $e = [v, w]$ in $L_v^{8k}$ such that $w \not\in B_{8k}$)\\
\hspace*{3mm}5.4 \hspace*{9mm} {\bf if} ($wt'(e) \geq m_0$)\\
\hspace*{21mm} add $e$ to $E_R$ but only keep the $k(16k-1)$ heaviest edges in $E_R$; \\
\hspace*{3mm}6. \hspace*{1mm} let $G_R = (V_R, E_R)$, where $V_R$ is the set of vertices incident to edges in $E_R$. 
\medskip
\strut} $$\boxit{\box4}$$
 \vspace{-8mm}
\caption{Constructing the reduced subgraph $G_R$ of a weighted graph $G$} 
\label{alg2}
\end{figure}

\noindent Our algorithm that constructs the reduced subgraph $G_R$ of the graph $G$ is presented in Figure~\ref{alg2}. 

We first prove the correctness of the algorithm {\bf RSubG} given in Figure~\ref{alg2}.

\begin{lemma}
\label{lem330}
The algorithm {\bf RSubG} given in Figure~\ref{alg2} constructs the reduced subgraph $G_R$ of the weighted 
graph $G$.  

\begin{proof} We start with the following observation: 

\smallskip 

{\bf Claim 1.} If the bounding set $B_{8k}$ in step 1 of the algorithm {\bf RSubG} contains $8k$ vertices, then the 
edge set $E_R$ in step 3 contains more than $k(16k-1)$ edges. 

{\it Proof of Claim 1.} Under the condition of the claim, place the $8k$ large-vertices in $B_{8k}$ into an ordered list 
$B_{8k}' = (v_1, v_2, \ldots, v_{8k})$, where all $v_i$ are large-vertices in $G$ whose $e_{8k}$-values are among 
the $8k$ largest $e_{8k}$-values over all large-vertices in $G$, and the vertices in the list $B_{8k}'$ are sorted 
non-decreasingly in terms of their $e_{8k}$-values. For any vertex $v_i$ in the list $B_{8k}'$, let $e = [v_i, w]$ be an 
edge incident to $v_i$, where $w$ is either a vertex $v_j$ in the list $B_{8k}'$ with $j < i$ or a vertex not in the list 
$B_{8k}'$. By definition, we have $wt'(e_{8k}(w)) \leq wt'(e_{8k}(v_i))$. Therefore, if $e \in L_{v_i}^{8k}$, i.e., 
if $e$ is among the $8k$ heaviest edges incident to $v_i$, then $e$ is also among the $8k$ heaviest edges incident 
to $w$, i.e., $e \in L_w^{8k}$, which means that the edge $e$ is in the trimmed subgraph $G_T$. As a result, among 
the $8k$ heaviest edges incident to the vertex $v_i$ in the list $B_{8k}'$, only those that are between $v_i$ and $v_p$, 
where $v_p$ is a vertex in $B_{8k}'$ with $i <p$, can be missing in the trimmed subgraph $G_T$. Thus, there are at 
least $i$ edges incident to the vertex $v_i$ in the trimmed subgraph $G_T$, i.e., the degree of the vertex $v_i$ in the 
trimmed subgraph $G_T$ is at least $i$. Let $G_T^{8k}$ be the graph that consists of the edges that are both in the 
trimmed subgraph $G_T$ and in the set $\bigcup_{v \in B_{8k}} L_v^{8k}$, then the degree sum of the vertices 
in $G_T^{8k}$ is at least $\sum_{i=1}^{8k} i = 4k(8k+1)$, which implies that the number of edges in the graph 
$G_T^{8k}$ (i.e., the number of edges in the set $E_R$) is at least $4k(8k+1)/2 > k(16k-1)$. This completes the 
proof of the claim.

\smallskip

Claim 1 directly implies the following result: 

\smallskip

{\bf Claim 2.} If the condition $|E_R| < k(16-1)$ in step 4 of the algorithm {\bf RSubG} holds, then the graph 
constructed in step 6 is the reduced subgraph $G_R$ of the graph $G$.

{\it Proof of Claim 2.}  If $|E_R| < k(16-1)$ in step 4, then by Claim 1, the set $B_{8k}$ contains fewer than $8k$ 
vertices, which implies that {\it all} large-vertices of the graph $G$ are included in the set $B_{8k}$, and the set 
$E_R$ constructed in step 3 contains {\it all} edges in the trimmed subgraph $G_T$ that are incident to {\it any} 
large-vertices in $G$. Therefore, the only edges in $G_T$ that are missing in the set $E_R$ are the edges whose 
both ends are small-vertices in $G$, i.e., vertices that are not in the set $B_{8k}$ (note that these edges are all in 
the trimmed subgraph $G_T$). Now steps 4.1-4.2 go through exactly all these edges and, together with the edges 
of $G_T$ that are already in the set $E_R$ after step 3, record the (up to) $k(16k - 1)$ heaviest edges. By the 
definition, these are exactly the edges that make up the reduced subgraph $G_R$. This proves the claim. 

\smallskip

The remaining case is that the set $E_R$ contains at least $k(16k-1)$ edges after step 3. Note that in this case, there 
can be large-vertices that are not included in the bounding set $B_{8k}$. After step 5.1, the set $E_R$ contains exactly  
$k(16k-1)$ edges, which are the $k(16k-1)$ heaviest edges among all edges in $G_T$ that are incident to vertices 
in $B_{8k}$. By the definition of the reduced subgraph, the edges deleted from the set $E_R$ in step 5.1 cannot be 
in the reduced subgraph $G_R$. Therefore, all edges in $G_T$ that are in the set $\bigcup_{v \in B_{8k}} L_v^{8k}$ 
and can possibly be in the reduced subgraph $G_R$ are included in the set $E_R$ after step 5.1. As a result, the 
edges that can possibly be in the reduced subgraph and are not yet included in the set $E_R$ after step 5.1 are those 
whose both ends are not in the set $B_{8k}$. Steps 5.2-5.3 examine all these edges.

\smallskip

{\bf Claim 3.} If the edge $e = [v, w]$ in step 5.3 of the algorithm {\bf RSubG} satisfies $wt'(e) \geq m_0$, then 
the edge $e$ is in the trimmed subgraph $G_T$. 

{\it Proof of Claim 3.} Let $e_0$ be the edge in the edge set $\bigcup_{v \in B_{8k}} L_v^{8k}$ that has the minimum 
edge weight, in terms of the edge weight function $wt'(\;)$. By the definition, $e_0$ must be the $(8k)$-th heaviest 
edge incident to a vertex $v_i$ in $B_{8k}$. Thus, $wt'(e_0) = wt'(e_{8k}(v_i))$ is the $e_{8k}$-value of the vertex 
$v_i$ in $B_{8k}$. Since the set $E_R$ constructed in step 5.1 is a subset of the set $\bigcup_{v \in B_{8k}} L_v^{8k}$, 
we have $m_0 \geq wt'(e_{8k}(v_i))$. Now, for the edge $e = [v, w]$ in step 5.3, where both $v$ and $w$ are not 
in $B_{8k}$, by the definition of the set $B_{8k}$, we must have $wt'(e_{8k}(v)) \leq wt'(e_{8k}(v_i))$ and 
$wt'(e_{8k}(w)) \leq wt'(e_{8k}(v_i))$ (recall that the $e_{8k}$-value of a small-vertex is defined to be $-\infty$). 
Therefore, if the edge $e = [v, w]$ satisfies $wt'(e) \geq m_0$, then we must have $wt'(e) \geq wt'(e_{8k}(v))$ 
and $wt'(e) \geq wt'(e_{8k}(w))$, i.e., the edge $e$ must be in the set intersection $L_v^{8k} \cap L_w^{8k}$, thus, 
in the trimmed subgraph $G_T$. This completes the proof of the claim. 

\smallskip

The edge set $E_R$ after step 5.1 contains exactly $k(16k-1)$ edges. By Claim 3, only edges in the trimmed subgraph 
$G_T$ can be added to $E_R$, and the set $E_R$ always contains exactly $k(16k-1)$ edges in the trimmed subgraph 
$G_T$. 

\smallskip

{\bf Claim 4.} If the edge $e$ in step 5.3 of the algorithm {\bf RSubG} satisfies $wt'(e) < m_0$, then the edge $e$ 
cannot be in the reduced subgraph $G_R$. 

{\it Proof of Claim 4.} If the edge $e$ is not in the trimmed subgraph $G_T$, then of course $e$ cannot be in the 
reduced subgraph $G_R$. Now suppose that $e$ is in the trimmed subgraph $G_T$. By the way the set $E_R$ is 
updated in step 5.4 and by Claim 3, the set $E_R$ always contains exactly $k(16k-1)$ edges in $G_T$ and the 
edge weight of any edge in $E_R$ is not smaller than $m_0$. Therefore, if $wt'(e) < m_0$, then the edge $e$ 
cannot be among the $k(16k-1)$ heaviest edges in the trimmed subgraph $G_T$, i.e., the edge $e$ is not in the 
reduced subgraph $G_R$. The claim is proved.

\smallskip

Therefore, if the edge set $E_R$ contains at least $k(16k-1)$ edges after step 3, which are the edges in both 
the trimmed subgraph $G_T$ and the set $\bigcup_{v \in B_{8k}} L_v^{8k}$, then step 5.1 deletes from the 
set $E_R$ some edges that obviously cannot be in the reduced subgraph $G_R$. Then, step 5.2-5.3 go through 
all edges that are not in the set $\bigcup_{v \in B_{8k}} L_v^{8k}$, ignore the edges that are obviously not in 
the reduced subgraph $G_R$ (Claim 4), and examine  the rest of the edges in the set in step 5.4 (by Claim 3, 
all edges examined in step 5.4 are in the trimmed subgraph $G_T$). As a consequence, all edges in the trimmed 
subgraph $G_T$ that are possibly in the reduced subgraph $G_R$ are examined in steps 5.1-5.4. Since we only 
keep the $k(16k-1)$ heaviest such edges, we conclude that after step 5, the set $E_R$ is the edge set of the 
reduced subgraph $G_R$. This gives us the following result:

\smallskip

{\bf Claim 5.} If the set $E_R$ contains at least $k(16k-1)$ edges after step 3 of the algorithm {\bf RSubG}, 
then the graph constructed in step 6 is the reduced subgraph $G_R$ of the graph $G$.

Combining Claim 2 and Claim 5 proves the lemma. 
\end{proof}
\end{lemma}

Now we can draw a conclusion for the algorithm {\bf RSubG} given in Figure~\ref{alg2}.

\begin{lemma}
\label{lem33}
There is an algorithm such that, for any $\epsilon > 0$, with probability at least $1 - \epsilon$, the algorithm on 
a weighted graph $G$ of size $N$ constructs the reduced subgraph $G_R$ of $G$ in time 
$O(N + k^2 + k \log(1/\epsilon))$ and space $O(k^2)$. 

\begin{proof}
By Lemma~\ref{lem330}, it suffices to verify that the algorithm {\bf RSubG} in Figure~\ref{alg2} satisfies the 
probability requirement and the time and space complexities stated in the lemma. 

To construct the bounding set $B_{8k}$ in step 1, we scan the graph $G$. For each large-vertex $v$, we construct  
the bounding list $L_v^{8k}$ as well as the $e_{8k}$-value for $v$. By Lemma~\ref{lem31}, this will take time 
$O(\mbox{deg}(v))$ and space $O(k)$, where the time complexity is, asymptotically, bounded by the amount of time 
for reading the edges incident to $v$. The $e_{8k}$-values of the large-vertices will be used as the keys in the 
construction of the bounding set $B_{8k}$. By Lemma~\ref{lem31}, with additional $O(L) = O(N)$ time and $O(k)$ 
space, where $L$ is the number of large-vertices in the graph $G$, we can construct the bounding set $B_{8k}$. 
Moreover, in this construction, we keep the bounding list $L_v^{8k}$ for at most $O(k)$ vertices. Since the bounding 
set $B_{8k}$ contains at most $8k$ vertices, the bounding set $B_{8k}$ and the bounding lists for the vertices in 
the set $B_{8k}$ can be constructed in time $O(N)$ and space $O(k^2)$  by step 1 of the algorithm. 

Step 2 of the algorithm constructs a hash function $H$ that is injective from the vertex set $B_{8k}$ to $[1 .. (8k)^2]$, 
where the set $B_{8k}$ contains at most $8k$ vertices. As we did for unweighted graphs in Theorem~\ref{theo1}, a 
hash function that maps the set of vertices in the graph $G$ to $[1 .. (8k)^2]$ and is randomly picked from a universal 
hashing class $\cal H$ has a probability at least $1/2$ to be injective from the set $B_{8k}$ to $[1 .. (8k)^2]$ \cite{clrs}. 
Therefore, with $\log(1/\epsilon)$-times of randomly picking a hash function from the universal hashing class $\cal H$, 
we will get a hashing function $H$ that is injective from the set $B_{8k}$ to $[1 .. (8k)^2]$, with probability at least 
$1 - \epsilon$. Note that with an initiated array of size $(8k)^2$, we can easily verify in time $O(k)$ if a given hash function is 
injective from $B_{8k}$ to $[1 .. (8k)^2]$. Therefore, in time $O(k \log(1/\epsilon) + k^2)$ and space $O(k^2)$, 
step 2 of the algorithm will construct the desired hash function $H$ with a probability at least $1 - \epsilon$. This is the 
only place in the algorithm where randomization is used.  

With the hash function $H$ constructed in step 2, we construct an array $B[1 .. (8k)^2]$ such that for each 
vertex $v$ in $B_{8k}$, the array element $B[H(v)]$ keeps the vertex $v$ as well as its $e_{8k}$-value. Now for any vertex 
$w$ in the graph $G$, we can test in constant time if $w$ is a vertex in the set $B_{8k}$, and in case it is, what 
is its $e_{8k}$-value. 

Recall that we have constructed the set $L_v^{8k}$ for each vertex $v$ in $B_{8k}$ in step 1. To construct the set 
$E_R$ in step 3, we need to identify the edges in these sets that are in the trimmed subgraph $G_T$. Let $e = [v, w]$ 
be an edge in the set $L_v^{8k}$ for a vertex $v$ in $B_{8k}$. If $w$ is not in $B_{8k}$, then since the $e_{8k}$-value 
of $w$ is smaller than that of $v$, the edge $e$ must be among the $8k$ heaviest edges incident to $w$. Thus, the edge 
$e$ must be in the graph $G_T$. On the other hand, if $w$ is in $B_{8k}$, then the edge $e$ is in $G_T$ if and only if 
$wt'(e)$ is not smaller than the $e_{8k}$-value of $w$. Thus, using the array $B[1 .. (8k)^2]$, we can test if the edge 
$e$ is in the trimmed subgraph $G_T$ in constant time. Finally, note that for an edge $e = [v, w]$ in $L_v^{8k}$ where 
$v \in B_{8k}$, if $w$ is not in $B_{8k}$, then the edge $e$ appears in the set $L_v^{8k}$ for exactly one vertex $v$ 
in $B_{8k}$, while if $w$ is in $B_{8k}$, then the edge $e$ appears in both $L_v^{8k}$ and $L_w^{8k}$. Therefore, 
for an edge $e = [v, w]$ with both $v$ and $w$ in $B_{8k}$, if we only consider the case when $v < w$, then we can 
avoid including multiple copies of an edge in the set $E_R$. Also note that the size of the set $E_R$ is bounded by that 
of $\bigcup_{v \in B_{8k}} L_v^{8k}$, which is $O(k^2)$. In conclusion, the set $E_R$ in step 3 can be constructed in 
time $O(k^2)$ and space $O(k^2)$. 

Steps 4-5 add new edges in the trimmed subgraph $G_T$ to the set $E_R$, and update the set $E_R$ so that the set 
$E_R$ only contains the $k(16k-1)$ heaviest edges seen so far. In order to keep the total processing time of steps 4-5 
to $O(N)$, we, instead of adding a new vertex directly to the set $E_R$, use a buffer of size $k^2$ to keep the new 
edges found in steps 4-5. Only after we collect $k^2$ new edges in the buffer, we combine these $k^2$ new edges with 
those in the set $E_R$, and select the $k(16k-1)$ heaviest to form the new set $E_R$. By Lemma~\ref{lem31}, this can 
be done in time $O(k^2)$ and space $O(k^2)$, contributing, in average, only constant time to each new edge. Also, to 
avoid including duplicated copies of an edge in the set $E_R$, for each edge $e = [v, w]$ encountered in steps 4-5 with 
$v \not\in B_{8k}$ and $w \not\in B_{8k}$, we only consider the edge when $v < w$. Putting all these together, we 
conclude that the total processing time of steps 4-5 is bounded by $O(N)$. The space complexity is $O(k^2)$.  

Summarizing the above discussions proves the lemma. 
\end{proof}
\end{lemma}

Now we return back to the p-WGM problem. Maximum matching on weighted graphs has been an extensively 
studied topic in theoretical computer science \cite{mmvc}. Currently, the best algorithm runs in time $O(n(m+n \log n))$ 
and space $O(m)$ on a weighted graph of $n$ vertices and $m$ edges \cite{gabow1990,gabow2018}, from which 
we can derive the following result. 

\begin{theorem}
\label{theo34}
There is an $O(k(m+n \log n))$-time and $O(m)$-space algorithm that on a weighted graph $G$ of $n$ vertices 
and $m$ edges, either constructs a maximum $k$-matching in $G$ or reports that no $k$-matching exists in $G$. 

\begin{proof}
This result is actually implied in the development of the $O(n(m+n \log n))$-time and $O(m)$-space algorithm due 
to Gabow \cite{gabow1990,gabow2018} that constructs a maximum matching in a weighted graph. In the following, 
we provide the necessary proofs for the parts that are not explicitly given in \cite{gabow1990,gabow2018} but are 
needed to achieve the stated result.

Let $G$ be a weighted graph. For a set $S$ of edges in $G$, we denote by $wt(S)$ the weight sum of the edges 
in $S$, and by $|S|$ the number of edges in $S$. Let $M$ be a matching in the graph $G$. Again we define an 
{\it augmenting path} relative to $M$ to be a simple path whose two ends are not covered by $M$ and whose 
edges go alternatively between edges not in $M$ and edges in $M$. The {\it weight-gain} of an augment path $P$ 
relative to the matching $M$ is defined to be $wt(P \setminus M) - wt(P \cap M)$. A {\it maximum augmenting path} 
relative to the matching $M$ is an augmenting path whose weight-gain is the largest over all augmenting paths relative 
to $M$. For a weighted graph $G$, we have the following (recall that for two sets $S_1$ and $S_2$, 
$S_1 \oplus S_2 = (S_1 \setminus S_2) \cup (S_2 \setminus S_1)$):

\smallskip

\noindent {\bf Claim.} Let $M_k$ be a maximum $k$-matching in the graph $G$, and let $P$ be a maximum 
augmenting path relative to $M_k$, then $P \oplus M_k$ is a maximum $(k+1)$-matching in the graph $G$.

\smallskip

For a proof of the claim, let $M_{k+1}$ be a maximum $(k+1)$-matching in the graph $G$. Then all connected 
components $C_1$, $C_2$, $\ldots$, $C_h$ of the graph $M_k \oplus M_{k+1}$ are either a simple cycle or a 
simple path. Since $|M_{k+1}| = |M_k| + 1$, at least one of the components of $M_k \oplus M_{k+1}$ is an 
augmenting path relative to $M_k$. Without loss of generality, assume that the component $C_h$ is an augmenting 
path relative to $M_k$, and let $C = C_1 \cup \cdots \cup C_{h-1}$. Then we have $|C \cap M_k| = |C \cap M_{k+1}|$. 
We claim that $wt(C \cap M_k) = wt(C \cap M_{k+1})$. In fact, if $wt(C \cap M_k) > wt(C \cap M_{k+1})$, then 
replacing the edges of the set $C \cap M_{k+1}$ in the $(k+1)$-matching $M_{k+1}$ with the edges of the set 
$C \cap M_k$ would give a $(k+1)$-matching whose weight is larger than that of $M_{k+1}$, contradicting the 
assumption that $M_{k+1}$ is a maximum $(k+1)$-matching. Similarly, if $wt(C \cap M_k) < wt(C \cap M_{k+1})$, 
then replacing the edges of the set $C \cap M_k$ in the $k$-matching $M_k$ with the edges of the set $C \cap M_{k+1}$ 
would give a $k$-matching whose weight is larger than that of $M_k$, contradicting the assumption that $M_k$ is a 
maximum $k$-matching. This equality $wt(C \cap M_k) = wt(C \cap M_{k+1})$ directly leads to the conclusion that 
the augmenting path $C_h$ relative to $M_k$ has its weight-gain equal to $wt(M_{k+1}) - wt(M_k)$. Since an augmenting 
path relative to $M_k$ with a weight-gain larger than $wt(M_{k+1}) - wt(M_k)$ would give a $(k+1)$-matching whose 
weight is larger than that of the maximum $(k+1)$-matching $M_{k+1}$, we conclude that the path $C_h$ is a 
maximum augmenting path relative to $M_k$, and augmenting the $k$-matching $M_k$ with the maximum 
augmenting path $C_h$ will result in the maximum $(k+1)$-matching $M_{k+1}$. This completes the proof of the claim.  

\smallskip

The algorithm given by Gabow \cite{gabow1990,gabow2018} is based on Edmonds' formulation of weighted 
matching as a linear program \cite{edmonds}. Starting with a maximum $0$-matching $M_0$ (i.e., an empty set), 
for each $i = 0, 1 \ldots$, the algorithm repeatedly finds a maximum augmenting path $P_i$ relative to the maximum 
$i$-matching $M_i$, and augments the matching $M_i$ along the path $P_i$ to obtain a maximum $(i+1)$-matching 
$M_{i+1}$ (whose correctness is given by the above claim). The process of finding a maximum augmenting path 
relative to a matching then augmenting the matching along the path is called a {\it phase}. Thus, after $k$ phases, 
a maximum $k$-matching is constructed for the graph $G$. On the other hand, if the process is stopped for a 
maximum $i$-matching $M_i$ with $i < k$ because there is no augmenting path relative to $M_i$, then we report 
that no $k$-matching exists in the graph $G$. Gabow \cite{gabow1990,gabow2018} has developed an algorithm 
that implements the computation of a phase in the above process in time $O(m + n \log n)$ and space $O(m)$. 
Combining these two results gives the proof of the theorem.
\end{proof}
\end{theorem}

For an instance $(G, k)$ of the p-WGM problem, the reduced subgraph $G_R$ of the graph $G$ contains 
$O(k^2)$ edges, thus no more than $O(k^2)$ vertices. Therefore, applying Theorem~\ref{theo34} to the 
reduced subgraph $G_R$, we conclude that a maximum $k$-matching in the reduced subgraph $G_R$ can be 
constructed in time $O(k(k^2 + k^2 \log k)) = O(k^3 \log k)$ and space $O(k^2)$. Bringing this result into 
Lemma~\ref{lem33} and letting $\epsilon = 1/k^{k^2}$ give the following theorem. 

\begin{theorem}
\label{theo35}
There is an algorithm for the p-WGM problem such that on an input $(G, k)$ where $G$ is a weighted 
graph of size $N$, with probability $1 - 1/k^{k^2}$, and running time $O(N + k^3 \log k)$ and space 
$O(k^2)$, the algorithm either constructs a maximum $k$-matching in $G$ or reports that no $k$-matching 
exists in $G$.  
\end{theorem}

We may not expect a very significant improvement on the complexity bounds given in Theorem~\ref{theo35}, 
based on the current status of maximum matching algorithms for weighted graphs. Indeed, if we measure the 
complexity of the algorithms in terms of the number $n$ of vertices in the graph, then the best algorithm 
for constructing a maximum weighted matching in a weighted graph takes time $O(n^3)$ \cite{gabow1973}. 
Since a graph has to have at least $2k$ vertices in order to contain a $k$-matching, the best we may expect 
for our reduction algorithm is to reduce the input graph into a reduced graph $G_R'$ of at least $2k$ vertices. 
Now applying the algorithm in \cite{gabow1973} to the reduced graph $G_R'$ will take time at least $O(k^3)$, 
which would give an algorithm of time $O(N + k^3)$ for the p-WGM problem. We also remark that directly 
applying the algorithm of time $O(n^3)$ in \cite{gabow1973} to the reduced subgraph $G_R$ in Lemma~\ref{lem33} 
does not give a better bound: the reduced subgraph $G_R$ in Lemma~\ref{lem33} may have $\Omega(k^2)$ 
vertices. 

Again, there seem no known algorithms that are specifically for solving the p-WGM problem. Chitnis {\it et al.} 
\cite{soda2016} studied the p-WGM problem on the dynamic graph streaming model, and proposed two 
randomized algorithms. As our discussions on the algorithms in \cite{soda2016} for the p-UGM problem (see Section 
3), we may remove the intricate (and expensive) operations that deal with edge deletions in the algorithms given in 
\cite{soda2016}, so that the algorithms can be used for solving the p-WGM problem. With this simplification, 
in order to have a success probability $1 - \epsilon$, the first streaming algorithm proposed in \cite{soda2016} 
would have update time (i.e., the time between reading two consecutive elements in the input) at least 
$O(\log W \log(1/\epsilon))$ and use space $O(k^4 W \log(1/\epsilon))$, where $W$ is the number of different 
values in the edge weights. As a consequence, if we use this algorithm to solve the p-WGM problem, the 
algorithm runs in time at least $O(\log W \log(1/\epsilon) N + k^4 W\log(1/\epsilon))$ and uses space 
$O(k^4 W\log(1/\epsilon))$. If we use the second algorithm proposed in \cite{soda2016}, with the above 
simplification, to solve the p-WGM problem, we would get an algorithm with running time at least  
$O(N \log k \log W + k^2 W \log(1/\epsilon))$ and space $O(k^2 W \log(1/\epsilon))$. More seriously, the 
second algorithm requires that the input weighted graphs have no matching of size larger than $k$, which makes 
the algorithm to be applicable to a much restricted class of graphs.

\section{Conclusion and final remarks}

Motivated by the recent algorithmic research in massive data processing, we proposed a parameterized computational 
model whose complexity bounds are measured by both input size $N$ and a parameter $k$, where $N$ is supposed 
to be extremely large while the parameter $k$ is a measure for the power of local resources (i.e., computational time 
and space) that can be used to deal with the massive data. We have used classical problems in computational optimization, 
the graph matching problems on both unweighted and weighted graphs, as examples to show how our model is used 
in effectively dealing with classical computational problems in massive data processing. In particular, we show how we can 
spend a linear-time pre-processing on the massive input data, with limited local memory space, to reduce a problem 
instance to an instance that is manageable by the limited local resources. Moreover, we showed how the local resources 
can be effectively managed to achieve the best or nearly best possible usage. In particular, we have presented an 
algorithm that finds a $k$-matching in an unweighted graph of size $N$ in time $O(N + k^{2.5})$ and space $O(k^2)$, 
and an algorithm that constructs a maximum weighted $k$-matching in a weighted graph of size $N$ in time 
$O(N + k^3 \log k)$ and space $O(k^2)$.  

Our algorithms for the graph matching problems are randomized algorithms, with exponentially small error bounds. If we 
use a balanced search tree to support the search and insertion operations in our process of large-vertices, instead of using 
injective hash functions, then our randomized algorithms will become deterministic algorithms. However, in their deterministic 
versions, our algorithm for solving the p-UGM problem in Theorem~\ref{theo4} will run in time $O(N \log k + k^{2.5})$ 
and space $O(k^2)$, and our algorithm for solving the p-WGM problem in Theorem~\ref{theo35} will run in time 
$O(N \log k + k^3 \log k)$ and space $O(k^2)$. 

The computational model we studied in the current paper suggests reconsiderations for many computational problems, 
including many classical ones, in the framework of massive data processing where the inputs are supposed to have 
extremely large size. For example, for two given vertices $s$ and $t$ in a weighted graph of size $N$, can we construct 
an $st$-path of length bounded by $k$ whose weight is the minimum over all $st$-paths of length bounded by $k$ in 
time $O(N + f_1(k))$ and space $O(f_2(k))$, where $f_1(k)$ and $f_2(k)$ are functions of the parameter $k$? If the 
answer if yes, what is the best we can get for $f_1(k)$ and $f_2(k)$? Note that Thorup's linear-time algorithm \cite{thorup} 
for the single-source shortest path problem seems not directly applicable here because of the space complexity. 

A particular research area where our model can be investigated is kernelization algorithms in parameterized computation 
\cite{kernelization}. Instances of a parameterized problem $Q$ take the format $(x, k)$, where $k$ is the parameter. 
A {\it kernelization algorithm} for the problem $Q$ on an input $(x, k)$ produces an instance $(x', k')$ such that 
$(x, k)$ is a yes-instance of $Q$ if and only if $(x', k')$ is a yes-instance of $Q$, and that the size of $x'$ and the 
value of the new parameter $k'$ are both bounded by a function of the original parameter $k$ that is independent of 
the size of the original input $(x, k)$. Most proposed kernelization algorithms run in polynomial time and were developed 
without much consideration on the efficiency of the algorithms. Recently, there have been studies on linear-time kernelization 
algorithms \cite{rolf}. On the other hand, space complexity has rarely been considered in kernelization algorithms. Many 
kernelization algorithms, including those proposed in \cite{rolf}, are based on the techniques that remove or modify 
``obvious'' structures in the input, which, intrinsically, requires space for storing the input and recording the changes, 
leading to demand of a large amount of space, and in many cases also to demand of super-linear time. On the other hand, 
the approach of kernelization seems to fit very well in dealing with massive data, and provides reduction and preprocessing 
techniques to reduce problem instances of very large size to instances of much small (thus manageable) size. In particular, 
kernelization algorithms whose running time is linear or nearly linear in terms of the input size, with limited space, are very 
interesting in this direction of research. We have initialized this line of research and obtained some preliminary results.

\end{document}